\documentclass{aa}

\usepackage{graphics}

\newbox\grsign \setbox\grsign=\hbox{$>$}
\newdimen\grdimen \grdimen=\ht\grsign
\newbox\laxbox \newbox\gaxbox
\setbox\gaxbox=\hbox{\raise.5ex\hbox{$>$}\llap
     {\lower.5ex\hbox{$\sim$}}}\ht1=\grdimen\dp1=0pt
\setbox\laxbox=\hbox{\raise.5ex\hbox{$<$}\llap
     {\lower.5ex\hbox{$\sim$}}}\ht2=\grdimen\dp2=0pt
\def\gax{\mathrel{\copy\gaxbox}}
\def\lax{\mathrel{\copy\laxbox}}

\begin{document}

\title{A High Resolution Spectroscopic Observation of CAL 83
with XMM-Newton/RGS
\thanks{Based on observations
obtained with XMM-Newton, an ESA science mission with instruments
and contributions directly funded by ESA Member States and the
USA (NASA).}
}

\author{
Frits Paerels \inst{1}, Andrew P. Rasmussen \inst{1}, 
H. W. Hartmann \inst{2}, J. Heise \inst{2}, 
A. C. Brinkman \inst{2}, 
C. P. de Vries \inst{2}, \\
and J.-W. den Herder \inst{2}
}

\offprints{F. Paerels}

\institute{
Columbia Astrophysics Laboratory, Columbia University,
550 West 120th St., New York, NY 10027, USA
\and 
SRON Laboratory for Space Research, Sorbonnelaan 2, 
3584 CA Utrecht, the Netherlands
}

\authorrunning{
F. Paerels et al.}

\titlerunning{
A High Resolution Spectroscopic Observation of CAL 83
with XMM-Newton/RGS}

\date{Received 2 October 2000/ Accepted 27 October 2000}

\abstract{
We present the first high resolution photospheric
X-ray spectrum of a Supersoft
X-ray Source, the famous CAL~83 in the Large Magellanic Cloud. The
spectrum was obtained with the Reflection Grating Spectrometer on
{\it XMM-Newton} during the Calibration/Performance Verification
phase of the observatory.
The spectrum 
covers the range 20--40 \AA\ at an approximately constant
resolution of 
0.05 \AA, and shows very significant, intricate detail, that is very
sensitive to the physical properties of the object. We present the
results of an initial investigation of the spectrum, from which we
draw the conclusion that the spectral structure is probably dominated
by numerous absorption features due to transitions in the L-shells of
the mid-$Z$ elements and the M-shell of Fe, 
in addition to a few strong K-shell features due
to CNO. 
}

\maketitle

\section{Introduction}

Supersoft X-ray Sources (or Supersoft Sources, SSS) were among the
first new discoveries made with {\it ROSAT}---in the first PSPC images
of the LMC, the SSS stood out immediately (Tr\"umper et al. 1991).
In retrospect, some
of these objects had been seen before with {\it Einstein}, but the
{\it ROSAT} detection in short order of several SSS
focused attention on them as a class.

It is now generally accepted that the SSS are white dwarf stars
undergoing steady nuclear burning in their envelopes. 
The inferred size of the radiating surface area, coupled with the
characteristically very soft optically thick spectra, point to this
idea as the natural interpretation for the observed emission.
Some SSS are
central stars of Planetary Nebulae, or related objects: shell-burning,
(mostly) single stars on their way to the white dwarf
graveyard.
The classical SSS, however, is
a high-luminosity ($L \gax 0.1~L_{\rm Edd}$), soft ($kT \lax 50$ eV)
low-mass binary with a white dwarf primary. In contrast to the case
for neutron
stars, for white dwarfs the efficiency of nuclear burning exceeds
the efficiency of gravitational energy conversion by accretion, by a
factor of order 100, so that SSS can sustain a high luminosity with a
sub-Eddington accretion rate. 

The basic model for the classical SSS has been worked out by van den
Heuvel et al. (1992). A relatively high-mass white dwarf (0.7--1.2
$M_{\sun}$) can sustain steady nuclear burning of accreted hydrogen
near its surface if the accretion rate is in the range 
$1-4 \times 10^{-7} M_{\sun}$ yr$^{-1}$. Such an accretion rate can
arise if the mass donor companion star is more massive than the white
dwarf (1.5--2.0 $M_{\sun}$). Mass transfer 
to the less massive star will cause the binary
orbit to shrink, and the donor will transfer mass on a rapid thermal
timescale. The predicted luminosities are in the range 
$6 \times 10^{36} - 1 \times 10^{38}$ erg s$^{-1}$, roughly
in the observed range for SSS. The binary evolution aspects of this
model, including an estimate for the birthrate and prevalence of SSS
binaries, have been addressed by Rappaport et al. (1994).

An important part of the interest in the SSS stems from the fact that
they may be a significant channel to SN Ia events. Mass is
deposited onto the white dwarf at such a high rate that the accreted
matter can burn non-eruptively, which makes the white dwarf mass grow
(in contrast to low accretion-rate binaries, such as classical
novae, which expel most of the previously accreted matter in explosive
events), eventually reaching the Chandrasekhar mass. Rappaport et al.
(1994) estimate that the rate of Galactic 
SN Ia supernovae associated with 
SSS binary evolution could be as high as $\sim 0.006$ yr$^{-1}$.

\begin{figure*}
 \resizebox{\hsize}{!}{\includegraphics{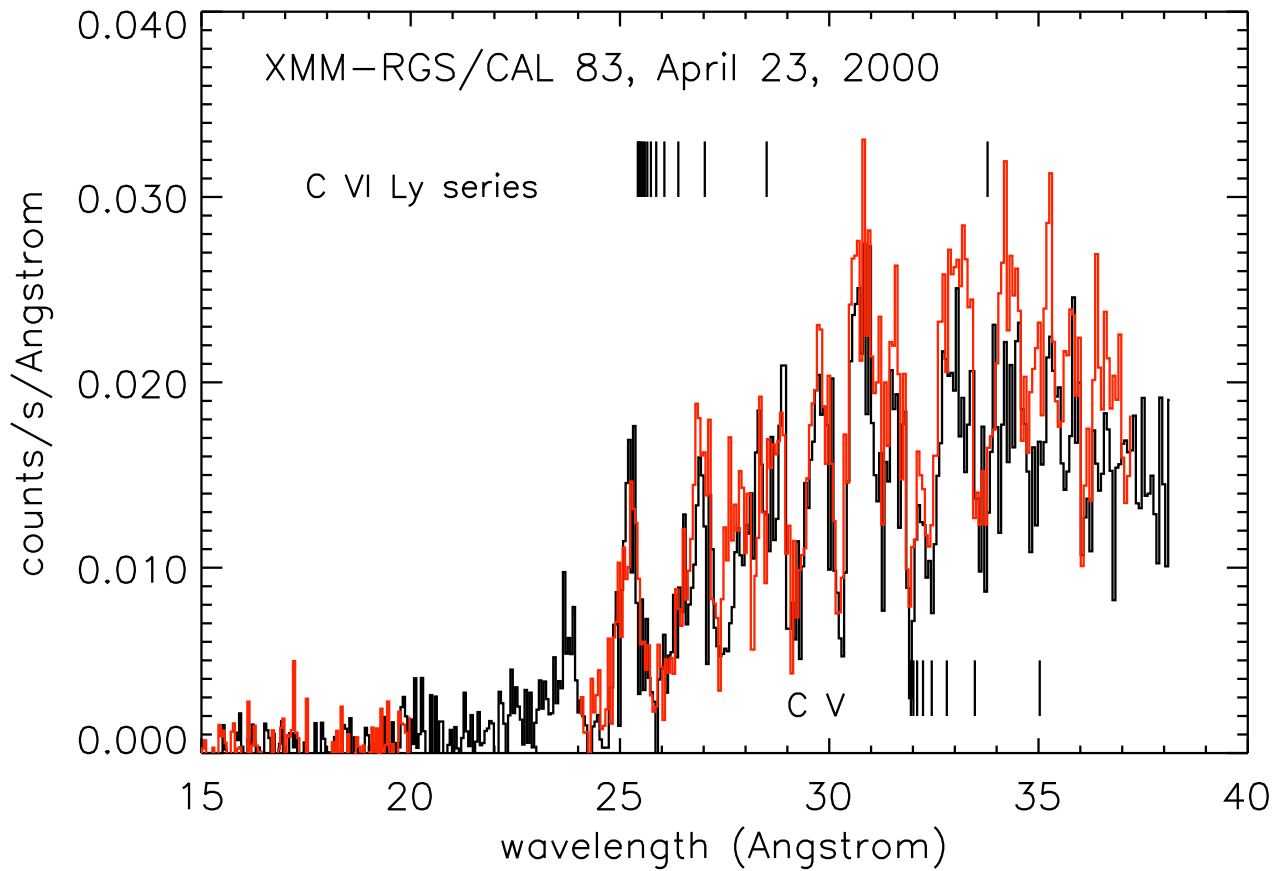}}
 \caption{
  Exposure corrected, background subtracted spectrum of CAL 83. We
  show the spectrum measured with RGS 1 (black histogram) overlaid on
  the spectrum measured with RGS 2 (red histogram), to emphasize the
  correspondence of coherent features in the spectrum in both
  spectrometer data sets. The wavelength bin equals 5 original RGS 81
  $\mu$m CCD spatial bins, which corresponds to approximately 0.06
  \AA, slightly larger than one resolution element. The range 
  $20.1-23.9$
  \AA\ is not present in the data from RGS2, because of malfunction of
  the drive electronics for one CCD chip in this spectrometer.
  Included for reference are wavelengths of
  the strongest 
  resonance transitions in hydrogen- and helium-like C.
  }
\end{figure*}

A high resolution soft X-ray spectroscopic observation can in
principle directly and definitively settle any remaining doubts about
the nature of the emission and the compact object, by detecting
characteristic photospheric structure in the spectrum. With a high
resolution spectrum, we can attempt to obtain an accurate estimate for
the fundamental stellar and binary parameters (luminosity, effective
temperature, radius and mass of the white dwarf, mass transfer rate),
which has a direct bearing on the evolutionary status and  ultimate
fate of the binary. In this {\it Letter}, we present preliminary
results of a high resolution spectroscopic observation of the
important SSS CAL 83 
(Cowley et al. 1984, Pakull et al. 1985, Crampton et al. 1987, Smale
et al. 1988, Parmar et al. 1998)
which directly addresses the above
issues.

\section{Data Analysis}

CAL 83 was observed by {\it XMM-Newton} (Jansen et al. 2000)
during the Calibration/Performance Verification phase
of the observatory, on April 23, 2000, for a total of 45.1 ksec. Data
was obtained with both the Reflection Grating Spectrometers (RGS,
den Herder et al. 2000) and the EPIC focal plane imaging cameras (Turner
et al. 2000). For the purpose of this paper, the high resolution
spectrum obtained with RGS holds the primary interest, and we will only
discuss the RGS data. 

The data were processed using custom software originally developed for
the analysis of RGS ground calibration data, which is nearly identical
in function to the RGS branch of the Science Analysis System
(SAS). Telemetered CCD events are read in frame by frame
and are offset--corrected on a pixel by pixel basis using 
median readout maps, compiled from about 40 diagnostic images
(den Herder et al. 2000) per CCD chip, those available from the ten most
recent revolutions (20 days). This process nearly eliminates
flickering pixels from the dataset. Remaining flickering and
warm pixels are removed from the datasets by thresholding duty cycle
maps (for each CCD), also generated from about 10 contemporaneous
revolutions.  Gain and CTI correction are performed to align the
signal/energy scale across all CCD readouts. Then, event
reconstruction is performed on a frame--by--frame basis where
connected pixels containing significant signal are recombined into
composite events and the composite event signal is calculated from the
sum of individual (corrected) signals.  

The standard event grade combination comprises events which fall within
a $2 \times 2$ pixel region, where two pixels diagonally opposed to
one another within the $2 \times 2$ region are considered two separate
events. The event coordinates are mapped into focal plane
two--dimensional angular coordinates (dispersion and
cross--dispersion). The dispersion coordinates are based on the known
geometry of the RGS combined with a preliminary in--flight calibration
based on the emission line spectra of \object{HR~1099} and
\object{Capella} which self--consistently determined the instrumental
boresight relative to the star tracker. The event positions are
corrected for small drifts in aspect.


Events are then windowed in the dispersion--pulseheight and focal
planes, using optimum extraction masks utilized by the response matrix
generator. 
Background subtraction was performed using background sampling regions
on either side (in the cross--dispersion direction) of the
source illuminated region. 
Finally, a spectrum file was created by histogramming the
corrected countrates in each dispersion channel. A response generator
was used to produce an observation specific response matrix that
provided an array of nominal wavelength/energy values corresponding to
each channel in the spectrum file. This procedure was performed for
both the
first and second orders for each spectrometer, but only the first
order spectra contained significant data. 

Figure 1 shows the exposure-corrected, background subtracted
spectrum from RGS 1 and 2. We show the spectrum obtained with the two
separate spectrometers superimposed, in order to emphasize that most
of the observed spectral detail is real, since it appears in both 
instruments. It is immediately obvious from a glance at Figure 1 that
blackbody spectral models, or indeed any smooth continuum spectrum,
is completely inadequate to characterize the spectrum of CAL 83, and,
by extrapolation, SSS in general. 
Previous low-resolution studies with {\it ROSAT} (Greiner et al. 1991;
see also Heise et al. 1994)
and {\it BeppoSAX} (Parmar et al.
1998) yielded circumstantial evidence that the spectrum had to 
deviate from a
simple blackbody (the bolometric correction associated with a 
blackbody yields 
estimates for the total luminosity that exceed the Eddington limit
by a large factor), but the spectral structure that must be present in
any real stellar photospheric spectrum is seen here for the first
time.

We have indicated the wavelengths of the strongest resonance lines in
the H- and He-like ions of C in Figure 1. At the approximate
densities and temperatures in the atmosphere (corresponding to $kT
\sim 50$ eV, $\log g \sim 8.5$), CNO are expected to be
primarily in their H- and He-like charge states, and one might expect
to see their absorption lines in the spectrum because of their
high abundances. 
Indeed, at
first sight, both C VI Ly$\alpha$ (33.73 \AA) and the C V $\lambda
34.97$ \AA\ $n=1-3$ resonance line appear to line up with a strong
absorption feature in the spectrum. Both the C VI Ly series limit, and
the C V ionization limit similarly coincide with a strong feature.
However, there is no such correspondence for any of the wavelengths in
N. Also, closer inspection reveals the presence of other deep
absorption features (e.g. the ones at 30.5 \AA, 29.0 \AA)
that do not correspond to any transitions in the H- and He-like ions
of C, N, or O.

\begin{figure}
 \resizebox{\hsize}{!}{\includegraphics{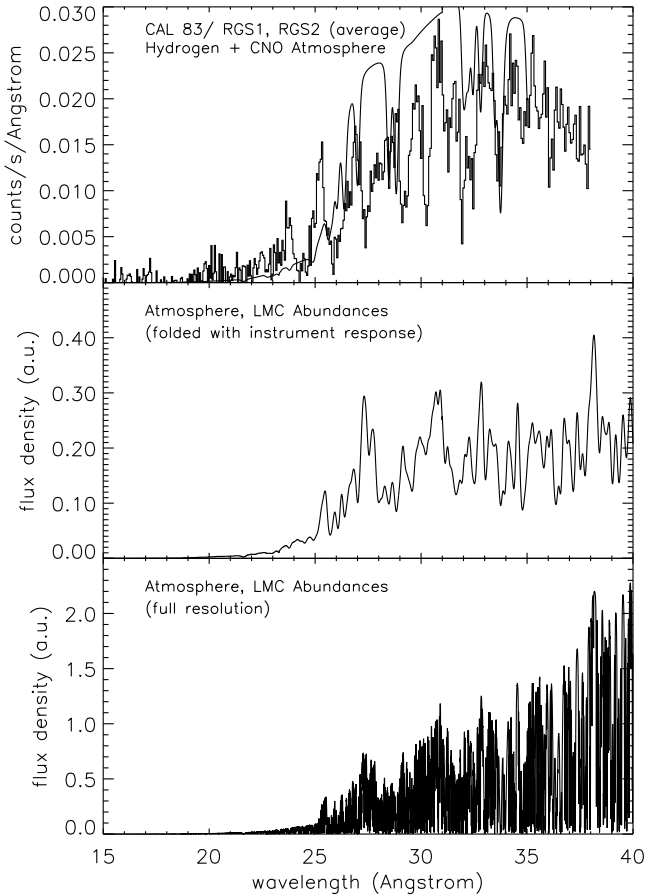}}
 \caption{
 {\it top panel:} Average spectrum of CAL 83 (average of RGS1 and 
 RGS2). Superimposed is a model atmospheres spectrum calculated for
 $kT_{\rm eff} = 45$ eV, $\log g = 8.5$, H + CNO (LMC abundances). The
 spectrum of CAL 83 clearly has a richer spectrum than this H+CNO
 atmosphere. {\it Middle panel}: Same atmosphere, with the abundant
 elements Ne-Ca and Fe added (LMC abundances); the spectrum has been
 convolved with the RGS instrument response. This spectrum
 qualitatively resembles the data. {\it Lower panel}: the same
 spectrum, but at the full resolution of the model calculation. The
 structure observed with the RGS is evidently not resolved.
  }
\end{figure}

We illustrate this finding in the top panel of Figure 2, which shows
the average of the two spectra from RGS 1 and 2, compared to a model
atmospheres spectrum calculated for a hydrogen atmosphere at $kT_{\rm
eff} = 45$ eV, $\log g = 8.5$, with trace amounts of CNO added at
their LMC abundance levels. This, and other models to be discussed in
this paper, have been calculated with Huben\'y's atmospheres code TLUSTY
(Huben\'y 1988; Huben\'y \& Lanz 1995), 
as applied to the conditions in white dwarfs (Hartmann
\& Heise 1997; Hartmann et al. 1999).
We added absorption by neutral gas, of column density
$N_{\rm H} = 6.5 \times 10^{20}$ cm$^{-2}$ (G\"ansicke et al. 1998), 
and convolved the model with the response of the RGS. Clearly, this
CNO dominated spectrum does not have the dense absorption structure we
see in CAL 83. 

In our next attempt at interpretation of the spectral structure, we
calculated a small grid of models around $kT_{\rm eff} = 45$ eV, $\log
g = 8.5$, this time with all abundant elements up to Fe, at their LMC
abundance levels. These spectra all indeed bear a qualitative
resemblance to the spectrum of CAL 83. An example (again at 
$kT_{\rm eff} = 45$ eV, $\log g = 8.5$) is shown in the middle panel
of Figure 2. The spectrum has been convolved with the instrument
response. The general structure of the model appears similar to the
spectrum of CAL 83. But most 
of the structure visible in this model is not
due to isolated, strong absorption lines, but instead is due to large
numbers of closely spaced, narrow absorption features from
transitions in the L shells of Ne through Ca, as well as
Fe M shell transitions. This dense forest of transitions is actually
not resolved by the RGS, as is illustrated by the bottom panel in
Figure 2, where we show the same model as in the middle panel, but
before convolution with the instrument response. 

We did not attempt a formal fit of these models to the measured
spectrum. Even though their qualitative appearance matches the shape
of the spectrum, they all differ from the data in significant details.
In a general sense, the fact that the discrete absorption spectrum has
not been resolved implies that we do not have the most powerful
spectroscopic diagnostics that would result from the presence of a few
isolated strong transitions in single ions. The dependence of the
shape and strength of these features on the properties of the
atmosphere (mainly density and temperature) would have given us 
unambigous estimates of the stellar parameters. Instead, the
sensitivity comes about through a dependence of the ionization balance
of the mid-$Z$ elements and Fe on temperature and density, and this
diagnostic is of course directly 
subject to uncertainties in the chemical composition of the
photosphere. The interpretation may also be sensitive to
the computational treatment of the problem and the completeness and
accuracy of the atomic structure data used to calculate the line
opacities, as well as to uncertainties in the treatment of collisional
broadening of the absorption lines.
We will address these issues in a future paper.


\section{Conclusions}

We have presented the first high resolution X-ray
spectrum of CAL 83. With {\it XMM-Newton} RGS, we finally
conclusively demonstrate the
photospheric origin of the soft X-ray emission. Moreover, our spectrum
qualitatively resembles the spectrum expected for the emission from a
very hot white dwarf star, and therefore confirms the basic model for
binary Supersoft X-ray Sources.

We detect radiation
from the photosphere between 20 and $\sim 40$ \AA, at an approximately
constant wavelength resolution of 0.05 \AA. The spectrum shows very
significant absorption structure. From comparison to model atmospheres
calculations at $kT_{\rm eff} \sim 45$ eV, $\log g \sim 8.5$, for
varying abundances, we conclude that most of this structure is due to
the superposition of numerous, unresolved narrow absorption features
due to transitions in the L shells of the mid-$Z$ elements, the M
shell of Fe, as well as in the K shell of C. A quantitatively reliable
estimate of the stellar parameters has to await a detailed analysis in
terms of a series of dedicated model atmospheres calculations.

\begin{acknowledgements}
The Columbia University group is supported by the US National
Aeronautics and Space Administration. The Laboratory for Space
Research Utrecht is financially supported by NWO, the Netherlands
Organization for Scientific Research.
\end{acknowledgements}

\end{document}